\documentclass[letterpaper,10pt]{article} 
\usepackage{graphicx}
\graphicspath{ {./Figures/} }
\usepackage{caption}
\usepackage{opticameet3} %% use version 3 for proper copyright statement

\newcommand\authormark[1]{\textsuperscript{#1}}

\usepackage{amsmath,amssymb}
\usepackage[colorlinks=true,bookmarks=false,citecolor=blue,urlcolor=blue]{hyperref} %pdflatex

\usepackage{tabularx}
\usepackage{float}

\begin{document}

\title{On the Required Secure Key Rate for Quantum-Secured 
Optical Channels}

% \author{Author name(s)}
% \address{Author affiliation and full address}
% \email{e-mail address}
%%Uncomment the following line to override copyright year from the default current year.
%\copyrightyear{2022}

\author{Farzam Toudeh-Fallah\authormark{1} , Robert Keys\authormark{1} , Dave Atkinson\authormark{1}}

\address{\authormark{1} Quantum Communication and Photonic Systems, Ciena Corporation\\
}

\begin{abstract}
The current maturity of the quantum-secured optical data channels based on the Quantum Key Distribution (QKD) technology is at the deployment level in metro environments, while R\&D efforts are also being conducted towards long-distance deployments. A great deal of research has been conducted on the achievable Secure Key Rate (SKR) for quantum channels. However, one of the major questions for network operators is the required SKR for the deployment of quantum-secured channels. This article addresses this question by defining the required SKR for quantum-secured optical channels and provides guidelines towards optimizing this parameter.
\end{abstract}

\section{Introduction}

The current key distribution methods utilize algorithms to encrypt the secret symmetric keys to be distributed amongst the communicating parties. These keys are then used to encrypt and decrypt the messages exchanged between the parties on the communication channels. However, these algorithm-based approaches will be vulnerable to attacks by quantum computers \cite{Mosca,Gidney}. In contrast, Quantum Key Distribution (QKD) relies on the principles of quantum mechanics to distribute quantum-secured secret keys, backed by theoretical proof of unconditional security\cite{Shor,Lo,Mayers}. Consequently, quantum-secured data channels based on the QKD technology have attracted a global interest. Instituting such channels over long-distances faces technological challenges and is the subject of world-wide research efforts. However, the maturity level of the short-distance quantum-secured channels is at a deployment level in metro environments\cite{Sandle}.

Secure Key Rate (SKR) (bits per second) is a measure of the key generation rate by QKD systems. Although several studies have been conducted on the achievable SKR for QKD systems \cite{Shields,Takeoka,Pirandola,Wang}, a major question raised by the network operators is the required SKR and how it relates to their specific operational environments. This article addresses this question with the intent of fostering a pragmatic approach towards defining this requirement for such deployments.

\section{Quantum-Secured Optical Channels}

\begin{figure}[H]
    \centering
    \includegraphics[width=\textwidth]{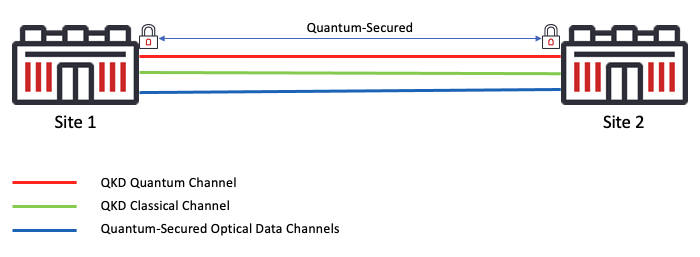}
    \caption{A generic model representing quantum-secured optical channels}
    \label{fig:Generic Model}
\end{figure}

Figure \ref{fig:Generic Model} depicts a generic model for establishing quantum-secured optical data channels.

As shown in this figure, QKD technology utilizes two channels (quantum and classical) to generate symmetric secret keys in both communication sites. These keys are then utilized to encrypt the optical data channels provisioned between the two sites using a symmetric encryption methodology such as AES-256, hence establishing quantum-secured optical channels. On some QKD systems, the quantum channel is established on the C band, in which case mostly a dedicated fiber will be required to prevent any interference, such as Raman Scattering, from the optical channels. However, some QKD systems establish the quantum channel on the O band, which allows the quantum channel to co-exist on the same fiber as the optical data channels. This study is independent of how quantum channel is established, its wavelength, encoding, protocol, or any other attributes.

\section{Defining the Required Secure Key Rate }

In the process of deploying quantum-secured optical channels, network operators must first establish the Required Secure Key Rate (RSKR). In high-capacity operational links, the entire C band, or in some cases, the entire C+L bands are filled with DWDM optical data channels and a deployment model must be able to quantum-secure the entire link. Therefore, the question arises on the Required SKR to satisfy this model. This requirement is dictated by the contributing factors on the optical data channels, as described in this section.

First, we define the Required Secure Key Rate Per Data Channel, which is the minimum key generation rate required for quantum-securing a data channel, designated as \(RSKR_{DC}\) (bits per second). This requirement depends on the secure key length (bits) and how fast the key is refreshed (per second) on each channel, as defined in equation \ref{eq:1}.

\begin{equation} \label{eq:1}
 RSKR_{DC} = L_k R_k
\end{equation}
where: \newline
\(RSKR_{DC}\): Required Secure Key Rate (RSKR) per data channel (bits/sec) \newline
\(L_k\): Required key length associated with data channel k (bits) \newline
\(R_k\): Required key refresh rate associated with data channel k (per second) \newline

A link usually consists of several multiplexed data channels (such as DWDM fiber links) and in order to calculate the Required Secure Key Rate for the entire link, the total number of data channels multiplexed on the link has to be taken into account as follows:

\begin{equation} \label{eq:2}
 RSKR_L = \sum_{k=1}^{N_{DC}}L_k R_k
\end{equation}
where: \newline
\(RSKR_L\): Required Secure Key Rate (RSKR) per link (bits/sec) \newline
\(N_{DC}\): Total number of channels per link \newline

Equation \ref{eq:2} is a general form applied to any link format with different key length and key refresh rate per channel, fix-grid or flex-grid, or any other format and provides the minimum SKR required for quantum-securing the entire operational band on a link. In the case of all channels on a link utilizing the same key length and key refresh rate, equation \ref{eq:2} turns into the following form:

\begin{equation} \label{eq:3}
 RSKR_L = L_k R_k N_{DC}
\end{equation}

Equation \ref{eq:3} clearly shows the dependency of the RSKR per link not only on the key length and the key refresh rate, but also on the number of data channels per link. Therefore, by reducing the number of data channels on a link, one can reduce the RSKR per link accordingly. Equation \ref{eq:3} could be applied to both fix-grid and flex-grid formatted links. However, to study an important consequence of this equation, let us consider the case of fix-grid DWDM-formatted links. In this case, equation \ref{eq:3} will result in:

\begin{equation}\label{eq:4}
 RSKR_L = L_K R_K \biggl\lfloor\frac{\delta_B}{\delta_C} \biggr\rfloor
\end{equation}
where: \newline
\(\delta_B\): Total spectrum of the operational band per link (Hz) \newline
\(\delta_C\): Spectral occupancy per data channel (Hz) \newline

Note that in equation \ref{eq:4} $\lfloor \hspace{0.1cm} \rfloor$ denotes the floor function.

\section{Impact of Optical Data Channel Configuration on the Link RSKR}

\begin{figure}[h]
    \centering
    \includegraphics[width=\textwidth]{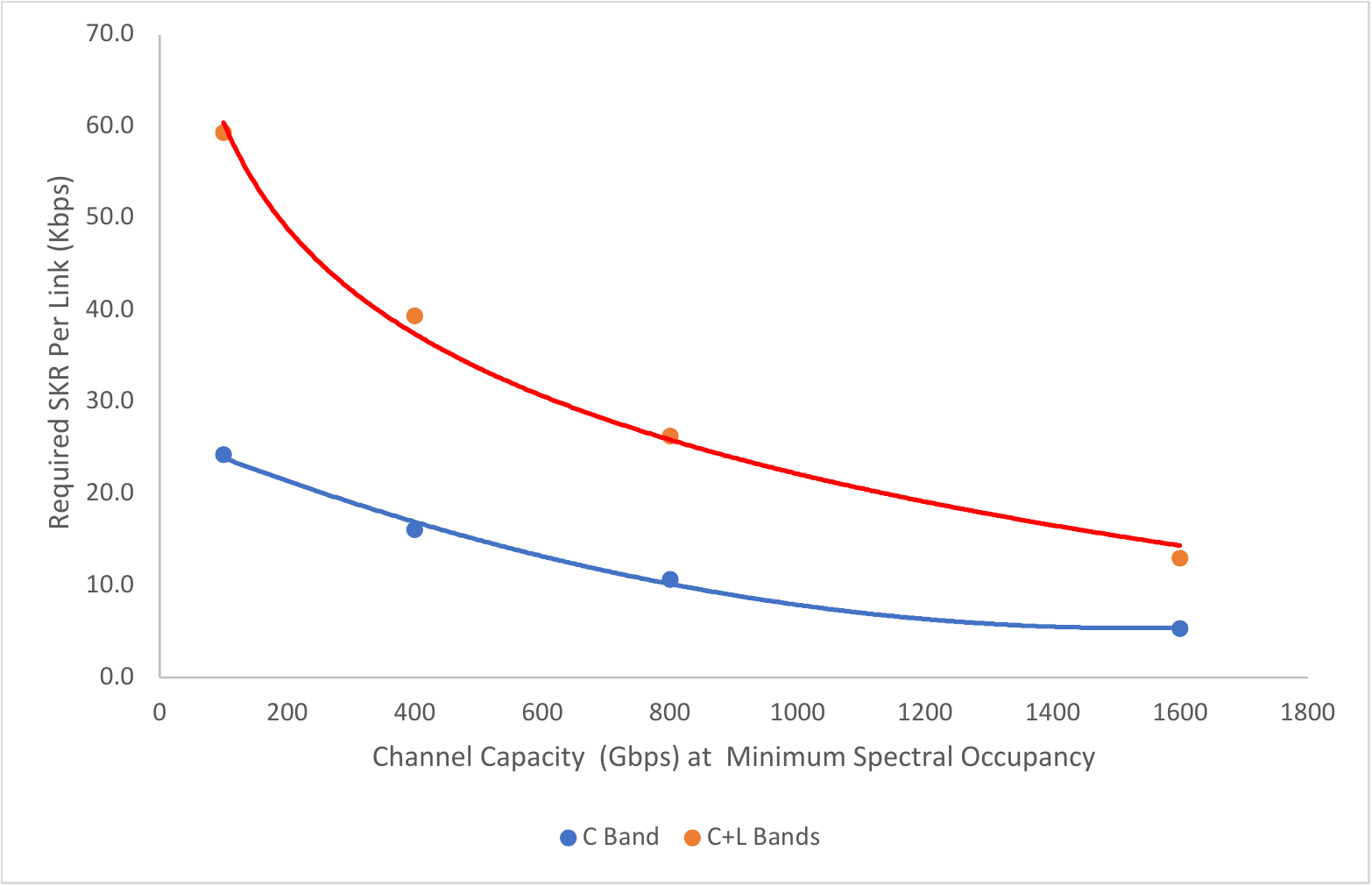}
    \caption{Required SKR Per Link vs Channel Capacity (at minimum spectral occupancy)}
    \label{fig:RSKR vs Channel Capacity}
\end{figure}

According to equation \ref{eq:4}, a shift to higher spectral occupancy per data channel $(\delta_C)$ would result into a drop in the RSKR per link.  An interesting case study is the increase in spectral occupancy towards achieving higher channel capacity. The relation between the channel capacity and spectral occupancy is a function of several factors, such as signaling rate and encoding mechanism. Based on equation \ref{eq:4}, Figure \ref{fig:RSKR vs Channel Capacity} shows the dependency between the RSKR per link and the channel capacity for C band-only and C+L band links. 

The channel capacity in Figure \ref{fig:RSKR vs Channel Capacity} is based on the minimum spectral occupancy and represents the experimental data on some of the current deployed optical data channels up to 800 Gbps and is extrapolated to higher channel capacities as specified in Table \ref{table:1}.

\begin{table}
\begin{center}
\begin{tabularx}{0.8\textwidth} { 
  | >{\centering\arraybackslash}X 
  | >{\centering\arraybackslash}X 
  | >{\centering\arraybackslash}X | }
 \hline
 Channel Capacity (Gbps) & Minimum Spectral Occupancy (GHz) \\
 \hline
 100  & 50 \\
\hline
 400  & 75 \\
\hline
800  & 112.5 \\
\hline
1600  & 225 \\
\hline
\end{tabularx}
\caption{Channel Capacity and the Corresponding Minimum Spectral Occupancy}
\label{table:1}
\end{center}
\end{table}

In addition, the following spectrum ranges were considered in Figure \ref{fig:RSKR vs Channel Capacity}:

C band: 191.35 - 196.10 THz

C+L band: 184.49 - 196.10 THz

It should be mentioned that Figure \ref{fig:RSKR vs Channel Capacity} is based on: 1) utilizing AES-256 as the encryption method for the optical data channels (at present the most widely used method in operational networks) and 2) a key refresh rate of one per second.

The analysis presented in Figure \ref{fig:RSKR vs Channel Capacity} reveals an important observation regarding the Required SKR per link. In both cases (C-band and C+L band), a decline in the Required SKR per link is observed as a function of increase in the optical data channel capacity (as it increases the spectral occupancy per channel). Therefore, according to this analysis, data aggregation to produce channels with higher throughput reduces the Require SKR per link and could be utilized as an optimization factor towards deployment of quantum-secured optical data channels. This correlation is more pronounced in the C+L band scenario.

It should be noted that the above-mentioned analysis is independent of the quantum channel used to generate the QKD keys. For instance, in addition to terrestrial quantum channels (established on any wavelength), this analysis is also applicable to the case where optical data channels are deployed on terrestrial fiber links, while the quantum channel is satellite-based (free-space in general).

Note that in addition to channel aggregation to higher capacity data channels, any other requirement on the optical data channels that causes an increase in the channel spectral occupancy will result in a decline in the Required SKR per link. Increasing the signaling rate towards achieving longer distance for a fixed channel capacity is another example.

\section{Conclusion}

In this article the Required Secure Key Rate (RKSR) was defined at both data channel and link levels for the deployment of quantum-secured optical channels. This is one of the fundamental requirements to be considered for such deployments. Upon defining the RSKR, the optical data channel spectral occupancy was identified as one of the most important contributing factors in determining this requirement. Accordingly, an RSKR optimization methodology based on data channel aggregation was discussed.

\section{Acknowledgements}

The authors would like to thank Charles Laperle and David Fasken for their review of this article and their comments.


\begin{thebibliography}{99} %% use BibTeX or add references manually


\bibitem{Mosca} M. Mosca and M. Piani, "Quantum Threat Timeline Report 2022", Global Risk Institute, December 2022

\bibitem{Gidney} C. Gidney and M. Ekera, "How to factor 2048 bit RSA integers in 8 hours using 20 million noisy qubits", Arxiv:1905.09749v3, 2021

\bibitem{Shor} P. Shor and J. Preskill, "Simple Proof of Security of the BB84 Quantum Key Distribution Protocol", Phys Rev Lett 85, 441, 2000

\bibitem{Lo} H.-K. Lo and H. Chau, "Unconditional Security of Quantum Key Distribution Over Arbitrarily Long Distances", Science, vol. 283, no. 5410, 1999

\bibitem{Mayers} D. Mayers, "Unconditional Security in Quantum Cryptography", Journal of the ACM, vol. 48, no. 3, 2001

\bibitem{Sandle} P. Sandle, "BT and Toshiba trial first commercial quantum-secured network", Reuters, 26 April 2022. [Online]. Available: https://www.reuters.com/business/bt-toshiba-trial-first-commercial-quantum-secured-network-2022-04-26/

\bibitem{Shields} A. Shields, "Performance Limits for Quantum Key Distribution Networks", ITU-T Workshop, Shanghai, 2019

\bibitem{Takeoka} M. Takeoka, S. Guha and M. Wilde, "Fundamental rate-loss tradeoff for optical quantum key distribution", Nature Communications 5, 5235, 2014. 

\bibitem{Pirandola} S. Pirandola, R. Laurenza, C. Ottaviani and L. Banchi, "Fundamental limits of repeaterless quantum communications", Nat Commun 8, 15043, 2017. 

\bibitem{Wang} X. Wang, S. Guo, P. Wang, W. Liu and Y. Li, "Realistic rate–distance limit of continuous-variable quantum key distribution", Optics Express, vol. 27, no. 9, 2019. 

\end{thebibliography}
\end{document}